\documentclass[a4paper,12pt]{article}

\usepackage{graphicx}
\usepackage{amsmath}

\begin{document}

\begin{center}

\textbf{\large{Magnetorotational Instability in Electrically Driven
Flow of Liquid Metal: Spectral Analysis}}\\
\vskip 0.5cm

\emph{I.V.Khalzov$^{1,2}$, A.I.Smolyakov$^{1,2}$, V.I.Ilgisonis$^1$}\\

\vskip 0.5cm

$^1$Russian Research Center "Kurchatov Institute", 1
Kurchatov Sq., Moscow, 123182, Russia\\
$^2$University of Saskatchewan, 116 Science Place, Saskatoon, SK,
S7N5E2, Canada
\end{center}

\begin{abstract}
The spectral stability of liquid metal differentially rotating in
transverse magnetic field is studied numerically by solving the
eigenvalue problem with rigid-wall boundary conditions. The
equilibrium velocity profile used in calculations corresponds to the
electrically driven flow in circular channel with the rotation law
$\Omega(r)\propto 1/r^2$. This type of flow profile is planned to be
used in new experimental device to test the magnetorotational
instability (MRI) in laboratory. Our analysis includes calculations
of the eigen-frequency spectra for both axisymmetric and
non-axisymmetric modes. It is found that for chosen device
parameters the flow is always spectrally unstable due to MRI with
the fastest growth rate corresponding to the axisymmetric mode.
\end{abstract}

\section{Introduction}

Magnetorotational instability (MRI) -- the instability of conducting
fluid, rotating in magnetic field -- is one of the most important
processes in astrophysics. MRI is believed to be responsible for
angular momentum transport in such objects as accretion disks formed
around massive black holes and active galactic nuclei. This
instability was originally discovered in 1959 by Velikhov
\cite{Vel}, who, though, did not consider it in an astrophysical
context. It was suggested only in 1991 \cite{Balbus} that a weak
magnetic field can linearly destabilize the otherwise stable
accretion disk thus directly leading to disk turbulence and orbital
angular momentum transport. MRI becomes operative when the angular
velocity decreases outwards, $d\Omega/dr<0$, which holds for
Keplerian flow in accretion disks ($\Omega\propto r^{-3/2}$).

Because of its importance in understanding of accretion disk
physics, a number of experiments have been proposed recently
\cite{Rud1}-\cite{Mariland} to study this instability in the
laboratory. In most of them a Couette flow of liquid metal between
two rotating cylinders placed in external magnetic field is supposed
to be used
However this simple design has its disadvantages. The main one is
that the equilibrium flow is controlled almost entirely by the
endcaps of the vessel -- so-called Ekman effect \cite{Holl}. In this
case the velocity profile is significantly different from Couette
flow, so the conditions for MRI may not be met.

Another way to rotate liquid metal is to apply radial electric
current between cylinders placed in vertical magnetic field, that
produces a Lorentz force driving metal. It is worthwhile noting that
in this case the effect of the Ekman layers at the top and bottom
insulating caps is considerably reduced. Actually one has to take
into account only Hartmann layers which are very small for large
Hartmann numbers and do not affect the main body of the flow. This
idea was proposed in Russian Research Center "Kurchatov Institute"
by E. P. Velikhov (2004, private communication) and the experimental
device is being designed there. A similar setup using the plasma
instead of liquid metal has been developed in Los Alamos
\cite{Los2}.

In this paper we present a linear stability analysis of the
electrically driven liquid metal flow in the circular channel for
the configuration relevant to Kurchatov experiment. It has been
shown in \cite{my} that in the stationary state the angular velocity
profile of such flow scales with radius as $\Omega\propto 1/r^2$
virtually throughout the entire channel cross section if the
Hartmann number is large enough (Ha$\gg1$). This dependence is a
particular case of the more general ones, $\Omega=a+b/r^2$ (Couette
flow) and $\Omega\propto r^{-q}$ (with $q=3/2$ corresponding to
Keplerian flow) whose stability has been studied analytically and
numerically in numerous papers.
Therefore some results obtained in those
papers  are relevant to our study. Let us briefly review them.

Originally MRI was discovered as a powerful local axisymmetric (with
azimuthal wave-number $m=0$) instability \cite{Vel, Balbus}.
Consideration of non-axisymmetric modes ($m\ne0$) in local approach
(for example, \cite{Ostr}) has shown that these modes become
stabilized with increasing values of $m$.

Importance of global modes and boundary conditions for MRI problem
have been first analyzed in \cite{Curry1}. In \cite{Curry2} authors
considered in details non-axisymmetric global modes with free
boundary conditions which are not appropriate for the system with
rigid walls.

A great deal of work in global stability study of magnetized Couette
flow (including our case with $a=0$) in the frame of non-ideal
magnetohydrodynamics (MHD) has been done by R\"{u}diger et al. (see,
for example, \cite{Rud1}, \cite{Rud3} and \cite{Rud2}). They found
marginal stability for different parameters of the system, but they
did not analyze the growth rate of MRI and its dependence on
azimuthal wave-number $m$.

The goal of this paper is the global spectral stability analysis of
the incompressible MHD flow with angular velocity profile
$\Omega\propto1/r^2$. The eigenvalue problem for this configuration
is set up in Section 2. The main body of results were obtained
numerically and are presented in Sections 3, 4, 5. In Section 3
axisymmetric modes ($m=0$) are considered and their classification
according to radial wave-number is made. Section 4 deals with
non-axisymmetric modes ($m\ne0$), the dependence of MRI growth rates
on azimuthal $m$ and axial $k$ wave-numbers is presented there. In
Section 5 we study marginal stability of the system under
consideration. In Section 6 we discuss the possibility of MRI in
electrically driven liquid metal flow.

\section{Statement of the Problem}

\begin{figure}[p]
\begin{center}
  \includegraphics[scale=0.3,angle=-90]{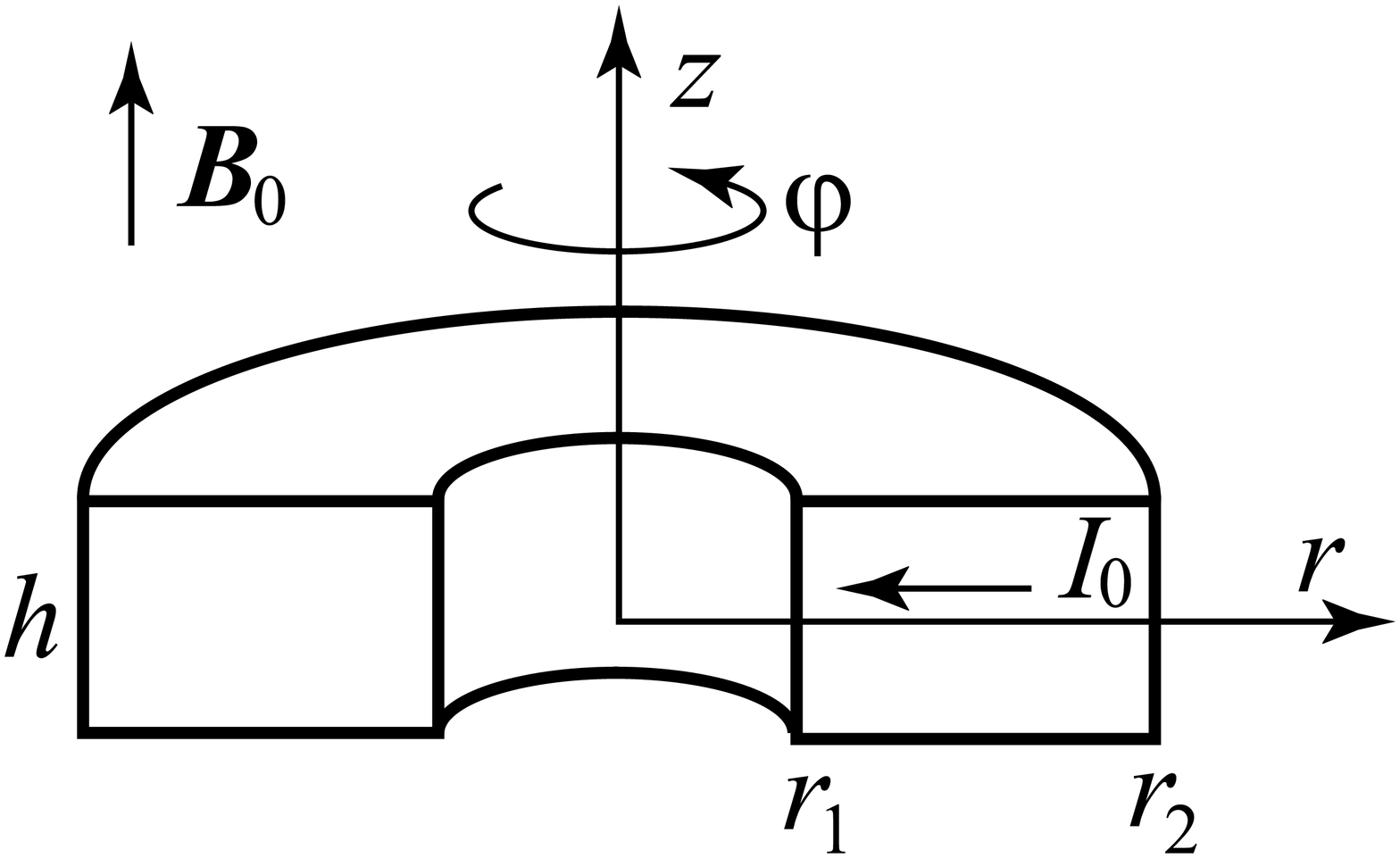}\\
  \caption{Geometry of the electrically driven flow for MRI study.\label{device}}
\end{center}
\end{figure}

We consider the following model of the experimental device proposed
in Kurchatov Institute (Fig. \ref{device}). A liquid metal (Sodium)
occupying a circular channel is driven by applying uniform vertical
magnetic field $\textbf{\emph{B}}_0$ and radial electric current
$I_0$. The stationary state of this configuration has been studied
in details in \cite{my}. It has been shown that in the cylindric
system of coordinates $(r,~\varphi,~z)$ the fluid rotates around the
$z$-axis with the angular velocity, which has the form
\begin{equation}
\label{Om} \Omega(r)=\frac{I_0}{4\pi r^2\sqrt{\sigma\rho\nu}}
\end{equation}
almost entirely in the channel if the Hartmann number is large
(Ha$\gg1$).

Despite the fact that the value of stationary velocity is determined
by the dissipative effects (conductivity $\sigma$ and viscosity
$\nu$ of the fluid)
in our study we neglect all of them and assume that the liquid metal
is a perfectly conducting, ideal, incompressible fluid with constant
density $\rho$. These assumptions are valid for the flow with high
Reynolds and magnetic Reynolds numbers -- such situation takes place
in the system under consideration. The model in this case is
simplified considerably without removal any essential physical
features. Also we note the paper \cite{PPPL3} where the calculations
of marginal stability for both viscous and inviscid fluids in the
case of Couette flow are presented  and they are shown to be almost
identical.


The fluid dynamics in our model is described by the equations of
ideal incompressible MHD:
\begin{equation}
\label{balance} \frac{\partial\textbf{\emph{V}}}{\partial
t}+(\textbf{\emph{V}}\boldsymbol{\cdot\nabla})\textbf{\emph{V}}=-\boldsymbol{\nabla}
P
+\frac{1}{4\pi\rho}(\textbf{\emph{B}}\boldsymbol{\cdot\nabla})\textbf{\emph{B}},
\end{equation}
\begin{equation}
\label{incompress} \boldsymbol{\nabla\cdot}\textbf{\emph{V}}=0,
\end{equation}
\begin{equation}
\label{frozen} \frac{\partial\textbf{\emph{B}}}{\partial
t}=\boldsymbol{\nabla\times}[\textbf{\emph{V}}\boldsymbol{\times}\textbf{\emph{B}}],
\end{equation}
\begin{equation}
\label{max} \boldsymbol{\nabla\cdot}\textbf{\emph{B}}=0,
\end{equation}
where
$P=(p+\textbf{\emph{B}}\boldsymbol{\cdot}\textbf{\emph{B}}/8\pi)/\rho$
is the normalized total pressure. In equilibrium state
($\partial/\partial t\to0$) the fluid, placed in uniform magnetic
field $\textbf{\emph{B}}_0=B_0\textbf{\emph{e}}_z$, rotates
differentially with velocity
$\textbf{\emph{V}}_0=r\Omega(r)\textbf{\emph{e}}_{\varphi}$. In
general, the function $\Omega(r)$ is arbitrary, since equilibrium
can always be maintained by a radial pressure gradient, but we will
consider here only the form of $\Omega(r)$ defined in equation
(\ref{Om}).

For the purpose of linear stability analysis, it is advantageous to
use a Lagrangian representation. We linearize the equations
(\ref{balance})-(\ref{max}) near the equilibrium state and introduce
the displacement vector $\boldsymbol{\xi}$, which is considered to
be a small quantity. The perturbations of velocity and magnetic
field are then expressed in linear approximation as follows
\cite{Frieman}:
\begin{eqnarray}
\label{dV}
\delta\textbf{\emph{V}}&=&\frac{\partial\boldsymbol{\xi}}{\partial
t} + (\textbf{\emph{V}}_0\boldsymbol{\cdot\nabla})\boldsymbol{\xi}
- (\boldsymbol{\xi\cdot\nabla})\textbf{\emph{V}}_0,\\
\label{dB}
\delta\textbf{\emph{B}}&=&\boldsymbol{\nabla\times}[\boldsymbol{\xi\times}
\textbf{\emph{B}}_0].
\end{eqnarray}
The linearized dynamics of $\boldsymbol{\xi}$ is obtained from
equation (\ref{balance}):
\begin{eqnarray}
\label{1} \frac{\partial^2\boldsymbol{\xi}}{\partial
t^2}&+&2(\textbf{\emph{V}}_0\boldsymbol{\cdot\nabla})\frac{\partial\boldsymbol{\xi}}{\partial
t} + (\textbf{\emph{V}}_0\boldsymbol{\cdot\nabla})^2\boldsymbol{\xi}
-
(\boldsymbol{\xi\cdot\nabla})(\textbf{\emph{V}}_0\boldsymbol{\cdot\nabla})\textbf{\emph{V}}_0=\\
&=& - \boldsymbol{\nabla}\delta P+
\frac{1}{4\pi\rho}(\delta\textbf{\emph{B}}\boldsymbol{\cdot\nabla})\textbf{\emph{B}}_0
+
\frac{1}{4\pi\rho}(\textbf{\emph{B}}_0\boldsymbol{\cdot\nabla})\delta\textbf{\emph{B}}.\nonumber
\end{eqnarray}
Besides, the equation (\ref{incompress}) gives:
\begin{equation}
\label{2} \boldsymbol{\nabla\cdot\xi}=0.
\end{equation}

The normal mode solutions to the system (\ref{1})-(\ref{2}) can be
sought in the form:
\begin{eqnarray}
\label{form1}
\boldsymbol{\xi}(r,\phi,z,t)&=&\boldsymbol{\xi}(r)e^{i(m\varphi+kz-\omega
t)},\\
\label{form2} \delta P (r,\phi,z,t)&=&\delta
P(r)e^{i(m\varphi+kz-\omega t)},
\end{eqnarray}
where $m$ (integer) and $k$ are the azimuthal and axial
wave-numbers, respectively, and $\omega$ is the eigen-frequency.
Since normal modes (\ref{form1}), (\ref{form2}) are generally
complex, to find the perturbations of physical quantities one has to
take either real or imaginary parts of righthand sides in equations
(\ref{dV}), (\ref{dB}) and (\ref{form2}).

Substituting the normal mode representation (\ref{form1}) into
equations (\ref{dV}) and (\ref{dB}) we find the perturbations:
\begin{eqnarray}
\label{dV2} \delta\textbf{\emph{V}}&=&-i\bar{\omega}\boldsymbol{\xi}
-
\frac{\partial\Omega}{\partial r}\,\xi_r\textbf{\emph{e}}_{\varphi},\\
\label{dB2} \delta\textbf{\emph{B}}&=&ikB_0\boldsymbol{\xi},
\end{eqnarray}
while the system (\ref{1})-(\ref{2}) can be rewritten as
\begin{eqnarray}
\label{r} &&(\omega_A^2-\bar{\omega}^2)\xi_r +
2i\Omega\bar{\omega}\xi_{\varphi} +
2\Omega\frac{\partial\Omega}{\partial r}\,\xi_r =
- \frac{\partial(\delta P)}{\partial r},\\
\label{phi} &&(\omega_A^2-\bar{\omega}^2)\xi_{\varphi} -
2i\Omega\bar{\omega}\xi_r
= - \frac{i m}{r}\delta P,\\
\label{z} &&(\omega_A^2-\bar{\omega}^2)\xi_z = - ik\delta P,\\
\label{div} &&\frac{1}{r}\frac{\partial(r\xi_r)}{\partial r}+\frac{i
m}{r}\xi_{\varphi}+i k\xi_z=0,
\end{eqnarray}
where we have introduced the shifted eigen-frequency,
$$
\bar{\omega}=\omega-m\Omega,
$$
and the Alfven frequency,
$$
\omega_A=kv_A=\frac{kB_0}{\sqrt{4\pi\rho}}.
$$
The analogous derivation for more general case including toroidal
magnetic field is given by \cite{Ogilvie}.

For computational convenience we introduce dimensionless quantities,
taking as a unit of length the radius of the inner cylindrical
boundary $r_1$ and as a unit of frequency the angular velocity at
this radius $\Omega_1$. Then we have:
$$
\tilde{r}=\frac{r}{r_1}, ~~~ \tilde{r}_1=1, ~~~
\tilde{r}_2=\frac{r_2}{r_1}, ~~~ \tilde{h}=\frac{h}{r_1}, ~~~
\tilde{k}=kr_1,
$$
$$
\tilde{v}_A=\frac{v_A}{r_1\Omega_1}, ~~~
\tilde{\omega}=\frac{\omega}{\Omega_1}, ~~~
\tilde{\omega}_A=\frac{\omega_A}{\Omega_1}=\tilde{k}\tilde{v}_A, ~~~
\tilde{\Omega}(\tilde{r})=\frac{1}{\tilde{r}^2},
$$
where $r_2$ is the radius of the outer cylindrical boundary.  In the
following consideration we will omit tildes to simplify notations.

The system (\ref{r})-(\ref{div}) can be reduced to one differential
equation which in non-dimensional terms is:
\begin{eqnarray}
\label{main} \left[\frac{(\bar{\omega}^2-\omega_A^2)x
u'}{m^2+k^2x}\right]' &+&
u\bigg[-\frac{\bar{\omega}^2-\omega_A^2}{4x} +
\frac{m\bar{\omega}}{m^2+k^2x}
\bigg(\Omega'-\frac{k^2\Omega}{m^2+k^2x} \bigg) +\nonumber \\
&+&
\frac{k^2\omega_A^2\Omega^2}{(m^2+k^2x)(\bar{\omega}^2-\omega_A^2)}
+ \frac{k^2\Omega(x\Omega'+\Omega)}{m^2+k^2x} \bigg]=0,
\end{eqnarray}
where $u=r\xi_r$, $x=r^2$ and the prime denotes the derivative with
respect to $x$. The further reduction of equation (\ref{main}) is
possible if the function $\Omega(r)$ of the form (\ref{Om}) is used.
In this case $\Omega\propto 1/r^2\propto1/x$ and the last term in
the equation is zero.

Since we consider the channel with rigid walls the velocity
components which are perpendicular to the boundary should vanish at
that boundary. It gives us the following boundary conditions
\begin{equation}
\label{BC} u(x_1)=u(x_2)=0,
\end{equation}
where $x_1$ and $x_2$ correspond to inner $r_1$ and outer $r_2$
cylindrical boundaries respectively. Besides, we have to quantize
axial wave-number $k$ appropriately, namely,
\begin{equation}
\label{k} k=\frac{\pi n_z}{h},
\end{equation}
where $n_z$ is integer and $h$ is the height of the channel.

The equation (\ref{main}) with boundary conditions (\ref{BC})
constitutes an eigenvalue problem. The primary objective of our
stability study is to determine the spectrum of eigen-frequencies
$\omega$ which has to be found as a part of the solution. Analyzing
this eigenvalue problem one can conclude that its solution (and
eigen-frequency spectrum as well) depends on five parameters: three
wave-numbers -- azimuthal $m$, axial $k$ and radial $n_r$ (it will
be introduced later), geometrical factor $x_2$ and dimensionless
Alfven velocity $v_A$. In this work we will focus on study of the
dependencies of eigen-frequencies $\omega$ on wave-numbers $m$, $k$
and $n_r$.

To solve the equation (\ref{main}) subject to the boundary
conditions (\ref{BC}) we used a standard shooting method adapted for
eigenvalue problem. For fixed values of problem parameters this
method allows one to find the eigen-functions $u(x)$ and
corresponding eigen-frequencies $\omega$. We have performed
calculations for the parameters of the experimental device proposed
in Kurchatov Institute (see Table \ref{t1}).
In numerical calculations the convenient normalization of
eigen-function $u'(x_1)=1$ has been chosen. The following sections
represent the results of our calculations.

\begin{table}[h]
\caption{Parameters of experimental device.\label{t1}}
\begin{tabular}{|l|l|l|} \hline\hline
Parameter & Value & Dimensionless Value \\
\hline
Inner radius, $r_1$ & 3 cm & 1 \\
Outer radius, $r_2$ & 15 cm & 5 \\
Height, $h$ & 6 cm & 2 \\
Velocity at inner radius, $v_1$ & 32 m s$^{-1}$ & 1 \\
Alfven velocity, $v_A$ & 8 m s$^{-1}$ & 0.25  \\
\hline
\end{tabular}
\end{table}

\section{Axisymmetric Perturbations, $m=0$}

In the axisymmetric case with $\Omega(x)=1/x$, equation (\ref{main})
is reduced to
\begin{equation}
\label{axisym} u'' + k^2u\bigg[-\frac{1}{4x} +
\frac{\omega_A^2}{x^3(\omega^2-\omega_A^2)^2} \bigg]=0.
\end{equation}
Multiplying this equation by complex conjugate $u^*$ and integrating
over the interval $[x_1,x_2]$ we obtain the following variational
form
\begin{equation}
\label{variat} \int\limits_{x_1}^{x_2} \bigg[-|u'|^2 -
\frac{k^2}{4x}|u|^2 +
\frac{k^2\omega_A^2}{x^3(\omega^2-\omega_A^2)^2}|u|^2 \bigg]dx=0.
\end{equation}
Some conclusions can be made from its consideration:
\begin{enumerate}
    \item The expression $(\omega^2-\omega_A^2)^2=\alpha^4$ is
    a real positive number. Therefore, $\omega^2=\omega_A^2\pm\alpha^2$ is
    real and eigen-frequencies $\omega$ are either pure real or pure imaginary.
    \item Eigen-frequency spectrum is symmetric with respect to
    both axis in the complex $\omega$-plane.
    \item Eigen-functions $u$ can always be chosen to be real.
    \item There is a degeneracy: each eigen-function $u$ corresponds to
    four eigen-frequencies $\omega$. Two of them are always real,
    $\omega=\pm\sqrt{\omega_A^2+\alpha^2}$, which means stability. Two
    others can be imaginary, $\omega=\pm\sqrt{\omega_A^2-\alpha^2}$,
    which means instability. This instability is known as the
    axisymmetric magnetorotational instability (MRI of axisymmetric modes).
\end{enumerate}

Besides the azimuthal and axial  wave-numbers, $m$ and $k$, to
describe eigen-frequencies completely we have to introduce the
radial wave-number $n_r$ which is defined in the case $m=0$ as a
number of zeros of corresponding real eigen-function. The
eigen-spectrum can be represented now by the formal disperse
relation $\omega=\omega(m,k,n_r)$ which relates each set of
wave-numbers $(m,k,n_r)$ with values of eigen-frequencies $\omega$.

For $m=0$ and lowest axial wave-number $k=\pi/2$ ($n_z=1$) the
spectrum of eigen-frequencies is shown in figure \ref{m=0,spec}.  As
one can see, there is a pure imaginary eigen-frequency in the upper
half-plane which corresponds to magneto-rotational instability. Also
the spectrum has two accumulation points at $\omega=\pm\omega_A$
(Alfven resonances). With approaching these points the radial
wave-number increases, i. e. the corresponding eigen-functions
become oscillating. In this limit the WKB approximation to the
equation (\ref{axisym}) works well (see, for example,
\cite{Curry1}).

Eigen-spectrum shown in figure \ref{m=0,spec} is formed from coupled
Alfven and slow magnetosonic modes. Unlike \cite{Goed} we did not
obtain the fast magnetosonic modes since our consideration is
restricted by incompressible MHD model.

It should be noted here that the Alfven resonance lines,
Re~$\omega=\pm\omega_A$, divide the complex $\omega$-plane into 3
regions: 1) Re~$\omega<-\omega_A$, 2)
$-\omega_A<$Re~$\omega<\omega_A$, and 3) Re~$\omega>\omega_A$. The
"outer" modes belonging to regions 1 and 3 are always stable (the
corresponding eigen-frequencies are real). The unstable modes can
emerge only in the central region 2 as a result of coupling of
originally stable "inner" modes. As we will show this tendency is
valid for non-axisymmetric modes as well.

In figure \ref{m=0,func} the samples of axisymmetric eigen-functions
with different radial wave-numbers $n_r$ are shown. No conclusion
about localization of these modes can be made in general.

\section{Non-Axisymmetric Perturbations, $m\ne0$}

In the case of non-axisymmetric perturbations one has to solve the
full equation (\ref{main}) in order to obtain the eigen-frequencies
and make a judgment on the stability of the system. Some important
properties of the eigen-frequency spectrum can be established prior
to solving the problem:
\begin{enumerate}
    \item The eigen-frequencies of non-axisymmetric modes may assume
    generally complex values.
    \item In the complex $\omega$-plane the eigen-frequency spectrum is symmetric with
    respect to the line Im~$\omega=0$, that is if $\omega$ is an
    eigen-frequency then its complex conjugate $\omega^*$ is also
    eigen-frequency. The presence of complex conjugate pair in the
    eigenvalue spectrum indicates the instability of the equilibrium
    under consideration since one of these eigen-frequencies corresponds
    to the perturbation growing in time.
    \item Eigen-functions $u$ are generally complex.
    \item There is no degeneracy in this case: each eigen-frequency
    corresponds to a different eigen-function.
\end{enumerate}

Unlike the axisymmetric case, in the case of $m\ne0$ the Alfven
resonances are not the lines any more. Instead they form two
"resonance" zones in the complex $\omega$-plane:
\begin{eqnarray}
\label{reg1} \min\limits_{x\in[x_1,~x_2]}(-\omega_A+m\Omega(x)) <
\textrm{Re}~\omega &<&
\max\limits_{x\in[x_1,~x_2]}(-\omega_A+m\Omega(x)),\\
\label{reg2}
\min\limits_{x\in[x_1,~x_2]}(\omega_A+m\Omega(x)) <
\textrm{Re}~\omega &<&
\max\limits_{x\in[x_1,~x_2]}(\omega_A+m\Omega(x)),
\end{eqnarray}
where the resonance condition Re~$\omega=\pm\omega_A+m\Omega(x)$ is
satisfied for some point $x\in[x_1,~x_2]$. As follows from our
numerical analysis for each set of wave-numbers $(m,k,n_r)$ there
are four different eigen-frequencies. Two of them are always real
(stable) and situated just outside the "resonance" zones, one in the
left outer region with
$$
\textrm{Re}~\omega <
\min\limits_{x\in[x_1,~x_2]}(-\omega_A+m\Omega(x)),
$$
another one in the right outer region with
$$
\textrm{Re}~\omega >
\max\limits_{x\in[x_1,~x_2]}(\omega_A+m\Omega(x)).
$$
The other two eigen-frequencies lie either in the "inner" region
between the resonances
$$
\max\limits_{x\in[x_1,~x_2]}(-\omega_A+m\Omega(x)) <
\textrm{Re}~\omega <
\min\limits_{x\in[x_1,~x_2]}(\omega_A+m\Omega(x)),
$$
or in the "resonance" zones if these zones are overlapped. In former
case the eigen-frequencies can be real or complex conjugate
depending on the values of problem parameters. In latter case the
eigen-frequencies are necessarily complex conjugate, that means the
presence of instability.

As a typical example of non-axisymmetric eigen-frequency spectrum
the spectrum for $m=1$ modes is shown in the figure \ref{m=1,spec}.
The resonance zones in this case are overlapped and all the "inner"
eigen-frequencies are complex conjugate (due to the resolution of
the figure \ref{m=1,spec} this fact is not so obvious for modes with
the small imaginary parts of eigen-frequencies). In figure
\ref{m=1,func} samples of these unstable eigen-modes are presented.

Note that in the case of unstable mode with very small imaginary
part of eigen-frequency a narrow singular layer is formed around the
resonance point $x_r$ defined by the condition
$$
\textrm{Re}~\omega=\pm\omega_A+m\Omega(x_r).
$$
The eigen-function has the steep gradients in the neighborhood of
this point (see Fig. \ref{m=1,func}b, \ref{m=1,func}c).

For the unstable non-axisymmetric mode the definition of its radial
wave-number $n_r$ is not so obvious as for axisymmetric one because
the non-axisymmetric eigen-function is complex. We resolve this
problem as follows: changing $m$ continuously from its initial value
to $m=0$ we find the corresponding axisymmetric mode with known
$n_r$, this $n_r$ is taken to be the radial wave-number of original
non-axisymmetric mode. Although such definition is mathematically
correct and non-ambiguous sometimes it leads to confusing results.
For example, comparing the figures \ref{m=1,func}b and
\ref{m=1,func}c one can see that both real and imaginary parts of
eigen-function with $n_r=2$ have less number of zeros than the
corresponding parts of eigen-function with $n_r=1$.

\begin{figure}[p]
\begin{center}
  \includegraphics[scale=0.4]{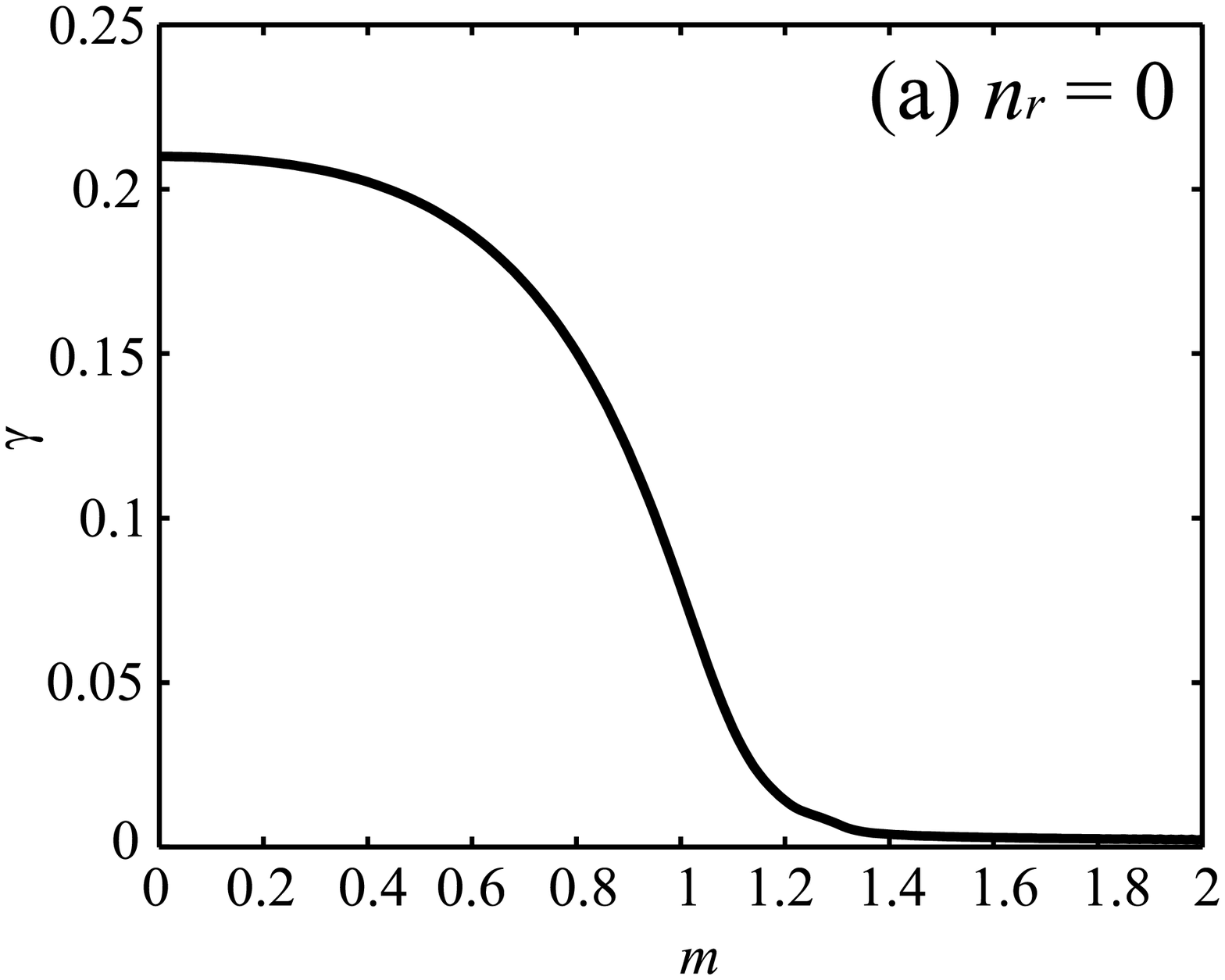}\\
  \includegraphics[scale=0.4]{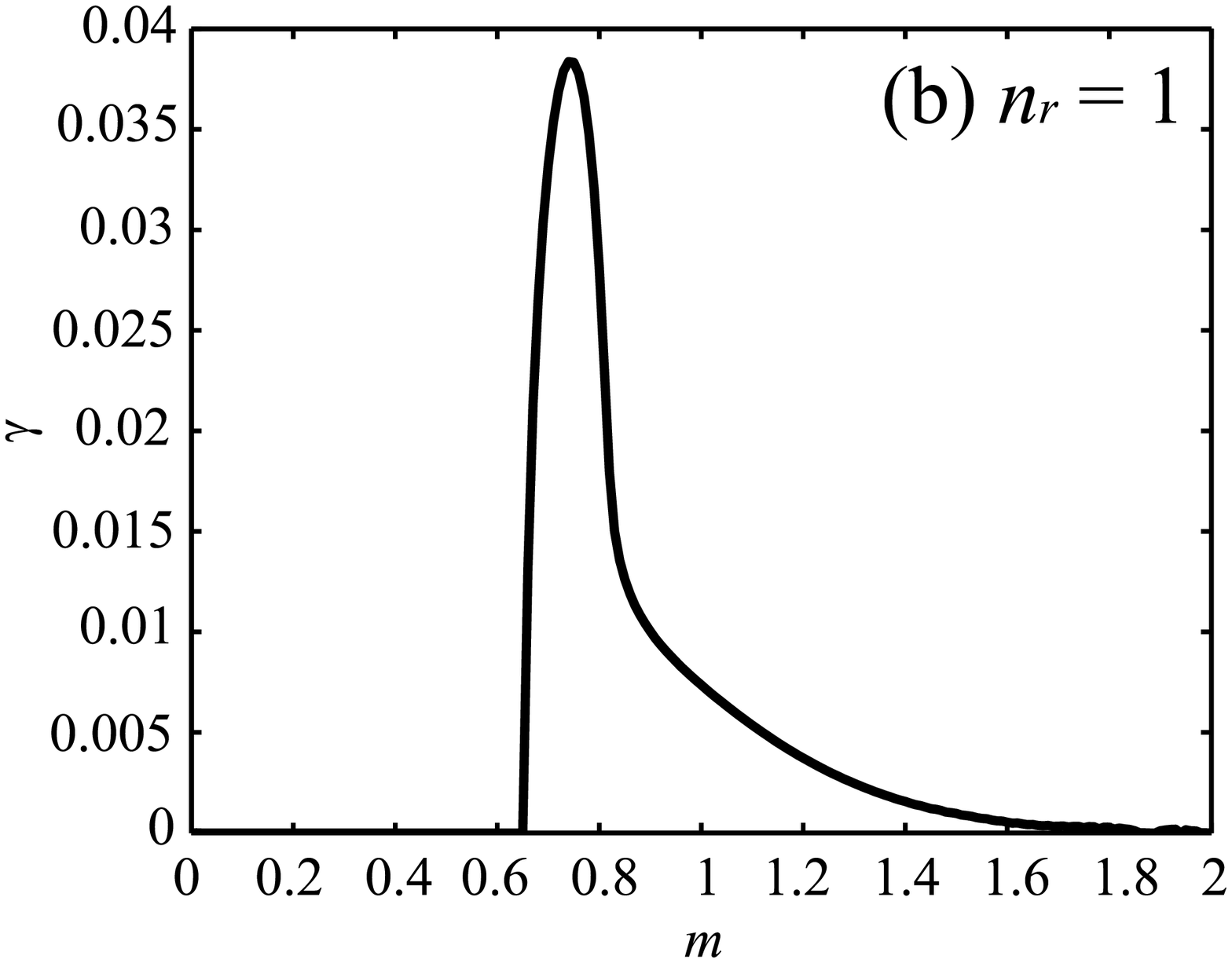}\\
  \includegraphics[scale=0.4]{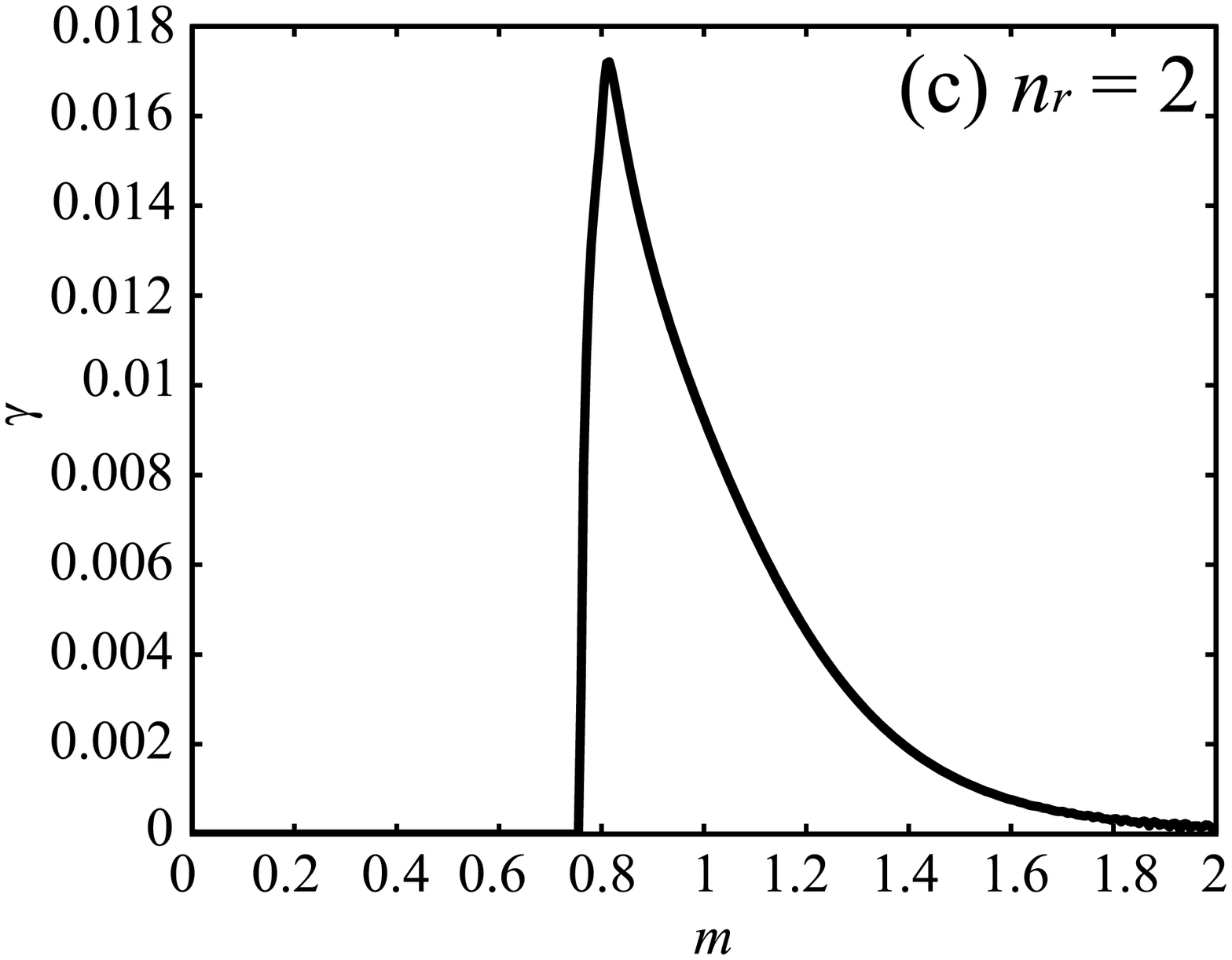}\\
  \caption{Growth rate dependence on azimuthal wave-number $m$ in
  case $k=\pi/2$: (a) $n_r=0$; (b) $n_r=1$; (c) $n_r=2$.\label{m=0-2}}
\end{center}
\end{figure}

Our main numerical results are related to the study of MRI growth
rate (or increment, $\gamma=\textrm{Im}~\omega$) as a function of
wave-numbers $m$, $k$, $n_r$. The figure \ref{m=0-2} demonstrate the
dependence of instability growth rate on the azimuthal wave-number
$m$ ($m$ is used here as continuous parameter) for eigen-modes with
different radial wave-numbers $n_r$. As one can see beginning with
$n_r=1$ the system under consideration has a threshold of
instability, $m_{cr}$, besides the value of $m_{cr}$ is larger for
larger $n_r$. If $m>m_{cr}$ the eigen-mode with corresponding radial
wave-number is always unstable, although the increment of
instability decreases with growth of $m$. This fact contradicts to
the local approach in which the unstable modes become stabilized
with increasing values of $m$ (see, for example, \cite{Ostr}). Also
it is worth noting that the maximum possible value of the increment
decreases with increase of radial wave-number $n_r$. In this sense
the most unstable mode is the mode with the lowest radial
wave-number, $n_r=0$.

\begin{figure}[p]
\begin{center}
  \includegraphics[scale=0.5]{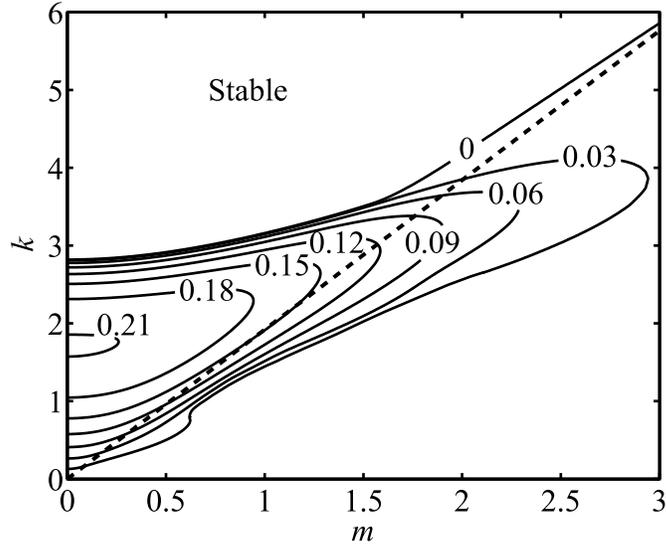}\\
  \caption{Growth rate dependence on azimuthal $m$ and axial $k$
  wave-numbers in case $n_r=0$. Solid contours show the levels of
  function $\gamma(m,k)$. The dashed line is the asymptote for the
  contour $\gamma(m,k)=0$.\label{m-k}}
\end{center}
\end{figure}

The figure \ref{m-k} demonstrates the dependence of the increment
$\gamma$ on the azimuthal $m$ and axial $k$ wave-numbers for modes
with $n_r=0$. According to this dependence, for every value of $m$
there exists a threshold value of axial wave-number $k_{cr}(m)$
after which the eigen-mode with $n_r=0$ is always stable. This
threshold value increases with growth of $m$ and, as we believe,
approaches the asymptote $k_{cr}\propto m$ (the explanation is in
the next section).

\begin{figure}[p]
\begin{center}
  \includegraphics[scale=0.5]{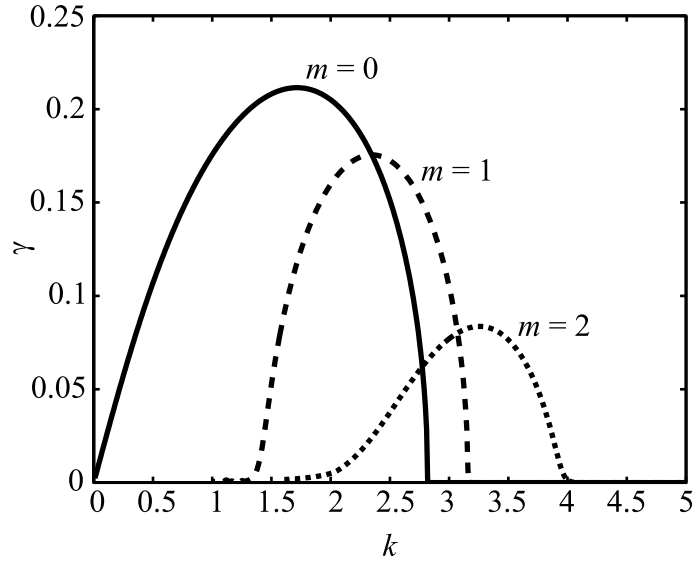}\\
  \caption{Growth rate dependence on axial wave-number $k$ for different $m$
  in case $n_r=0$. Solid line is for $m=0$, dashed -- for $m=1$,
  dotted -- for $m=2$.\label{dep_k}}
\end{center}
\end{figure}

Dependencies of the increment on $k$ for some $m$ are shown in more
details in figure \ref{dep_k}. In fact, this figure represents the
sections of the figure \ref{m-k} at corresponding $m$.

Since the minimal axial wave-number $k$ is determined by the height
of the channel (see eq. \ref{k}) we conclude that choosing the
height to be small enough (that is the minimal $k$ is bigger than
$k_{cr}$) one can suppress the instabilities with low $m$. In
particular, the situation is possible when there is instability for
$m=1$ mode but there is no instability for axisymmetric $m=0$ mode.

\section{Marginal Stability}

As we have mentioned above \emph{if the resonance zones in the
complex $\omega$-plane are overlapped all the inner modes become
unstable}. Although we checked this statement only for the small
range of $m$ (from 0 to 3) we believe that this is also true for
larger $m$. Our limitations in $m$ are related to the computer
accuracy: in fact the growth rates for large $m$ are so small that
they can not be resolved in the code even by double float
representation.

Accepting this statement as a hypothesis we obtain from conditions
(\ref{reg1}) and (\ref{reg2}) that all inner modes (with any radial
wave-number  $n_r$) become unstable if
$$
\max\limits_{x\in[x_1,~x_2]}(m\Omega(x)-\omega_A)=\min\limits_{x\in[x_1,~x_2]}(m\Omega(x)+\omega_A).
$$
Taking into account that $x_1=1$ and the angular velocity profile is
$\Omega(x)=1/x$, we obtain
$$
m-\omega_A=\frac{m}{x_2}+\omega_A,
$$
which gives a critical value of axial wave-number $k$ as a function
of $m$:
\begin{equation}
\label{cr}k_{cr}(m)=\frac{m}{2v_A}\bigg(1-\frac{1}{x_2}\bigg).
\end{equation}
It means that all "inner" modes (with any $n_r$) become unstable if
the axial wave-number $k$ is less than the critical value defined by
(\ref{cr}).

\begin{figure}[p]
\begin{center}
  \includegraphics[scale=0.5]{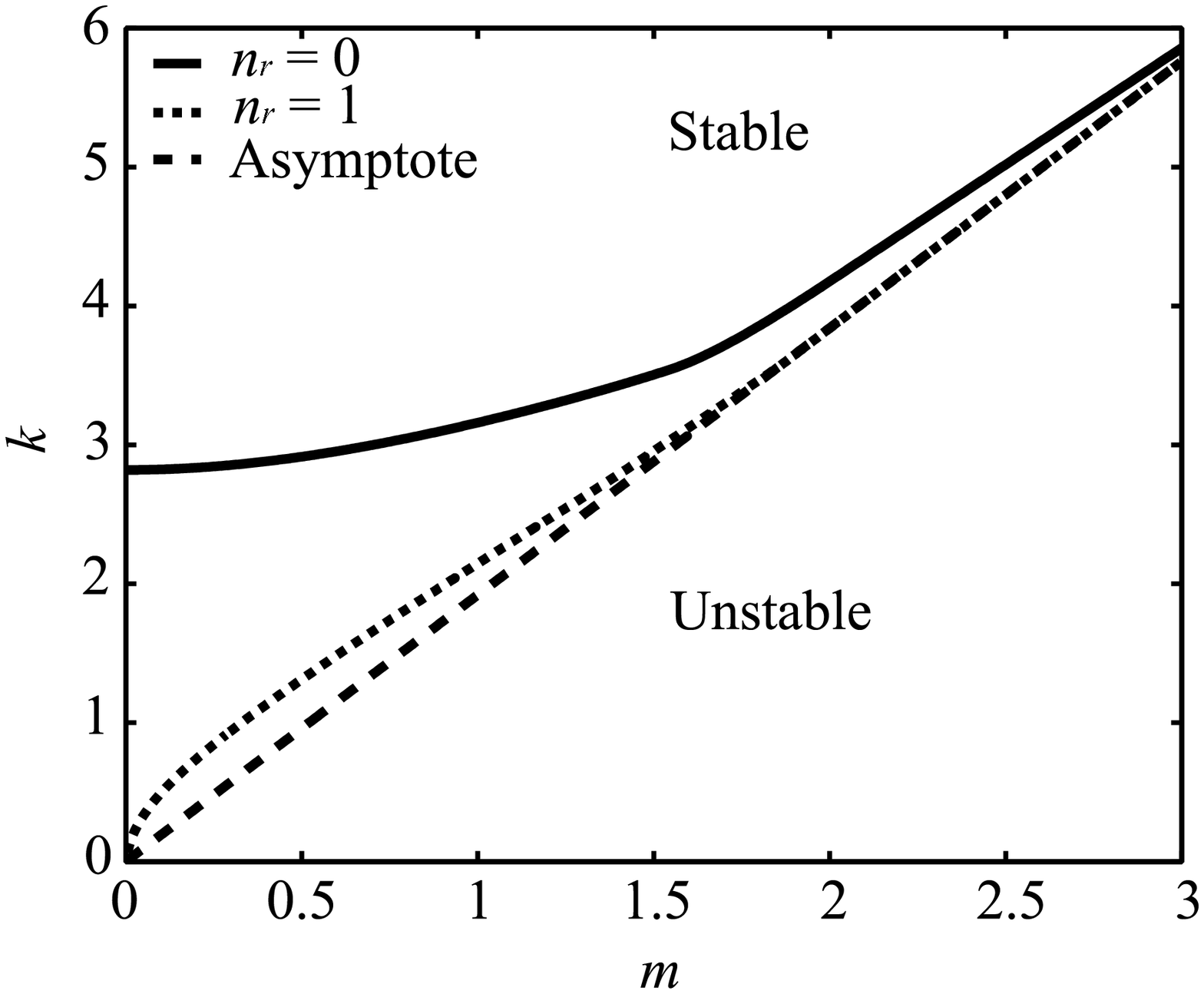}\\
  \caption{Marginal stability in $(k,m)$-plane for different radial wave-numbers
  $n_r=0$. Solid line is for $n_r=0$, dotted -- for $n_r=1$;
  dashed line is their asymptote.\label{marg}}
\end{center}
\end{figure}

In reality the instability can occur for larger values of $k$, the
actual marginal stability curves depend on the radial wave-number
$n_r$. In the figure \ref{marg} the marginal stability curves
calculated for $n_r=0$ and $n_r=1$ are shown. For comparison, the
critical line $k_{cr}(m)$ is also plotted. It is quite clear that
this line is asymptote for both presented curves in the limit
$m\to\infty$. Besides the critical line almost coincides with the
marginal stability in the case of large radial wave-numbers,
$n_r\gg1$.

Note that the system under consideration has the only unstable
axisymmetric mode with radial wave-number $n_r=0$. All axisymmetric
modes with $n_r\geq1$ are stable regardless of the value of $k$.
From the figure \ref{marg} and expression (\ref{cr}) it also follows
that in this system
there are always non-axisymmetric eigen-modes which are spectrally
unstable due to magnetorotational instability.

\section{Conclusions}

We have examined the spectral stability of electrically driven
liquid metal flow in the model relevant to the new MRI experiment
and presented a detailed numerical analysis of eigen-functions and
eigen-frequencies inherent in this model. The  obtained results show
that the rotation profile given by the expression (\ref{Om}) is
always unstable due to MRI. For the chosen parameters of the
experimental device the highest growth rate of instability $\gamma$
corresponds to the axisymmetric mode $m=0$ with the lowest radial
wave-number $n_r=0$. Consequently, in the experiment with
electrically driven flow one should expect the rapid excitation of
axisymmetric magnetorotational instability. The modes with other
wave-numbers are also presented but they are not so robust as the
axisymmetric one.

From our analysis it turns out that the growth rate of axisymmetric
MRI is quite sensitive to the device parameters. For example,
choosing the height of the channel to be small enough (to make the
smallest possible $k$ in the system bigger than its threshold value
for $m=0$) one can suppress the axisymmetric MRI while the unstable
modes with $m\ne0$ will still exist.

The model which we have considered here is, of course, idealized. We
did not take into account the dissipative effects (viscosity and
resistivity) as well as the deviations of the equilibrium velocity
from the profile given by the equation (\ref{Om}). These effects can
change the conditions for MRI excitation in the experiment; they
will be a subject of our future study.

We would like to acknowledge E. P. Velikhov for statement of this
problem and V. P. Lakhin for useful discussions. This work is
supported in part by NSERC Canada, NATO Collaboration Linkage Grant
and Global Partners Award (University of Saskatchewan, Canada).

\begin{figure}[p]
\begin{center}
  \includegraphics[scale=0.5]{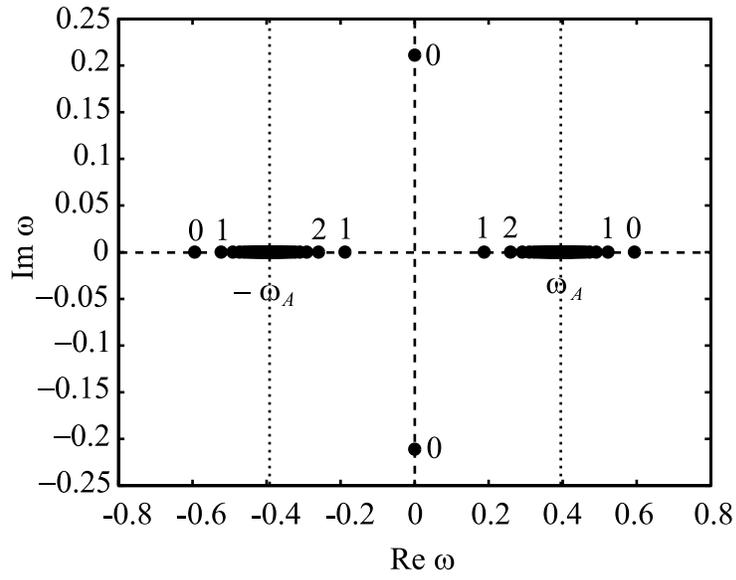}\\
  \caption{Eigen-frequency spectrum in case $m=0$, $k=\pi/2$. Dashed
  lines represent real and imaginary axes, dotted lines correspond to Alfven
  resonances. Radial wave-numbers $n_r$ are shown for some
  modes.\label{m=0,spec}}
\end{center}
\end{figure}

\begin{figure}[p]
\begin{center}
  \includegraphics[scale=0.4]{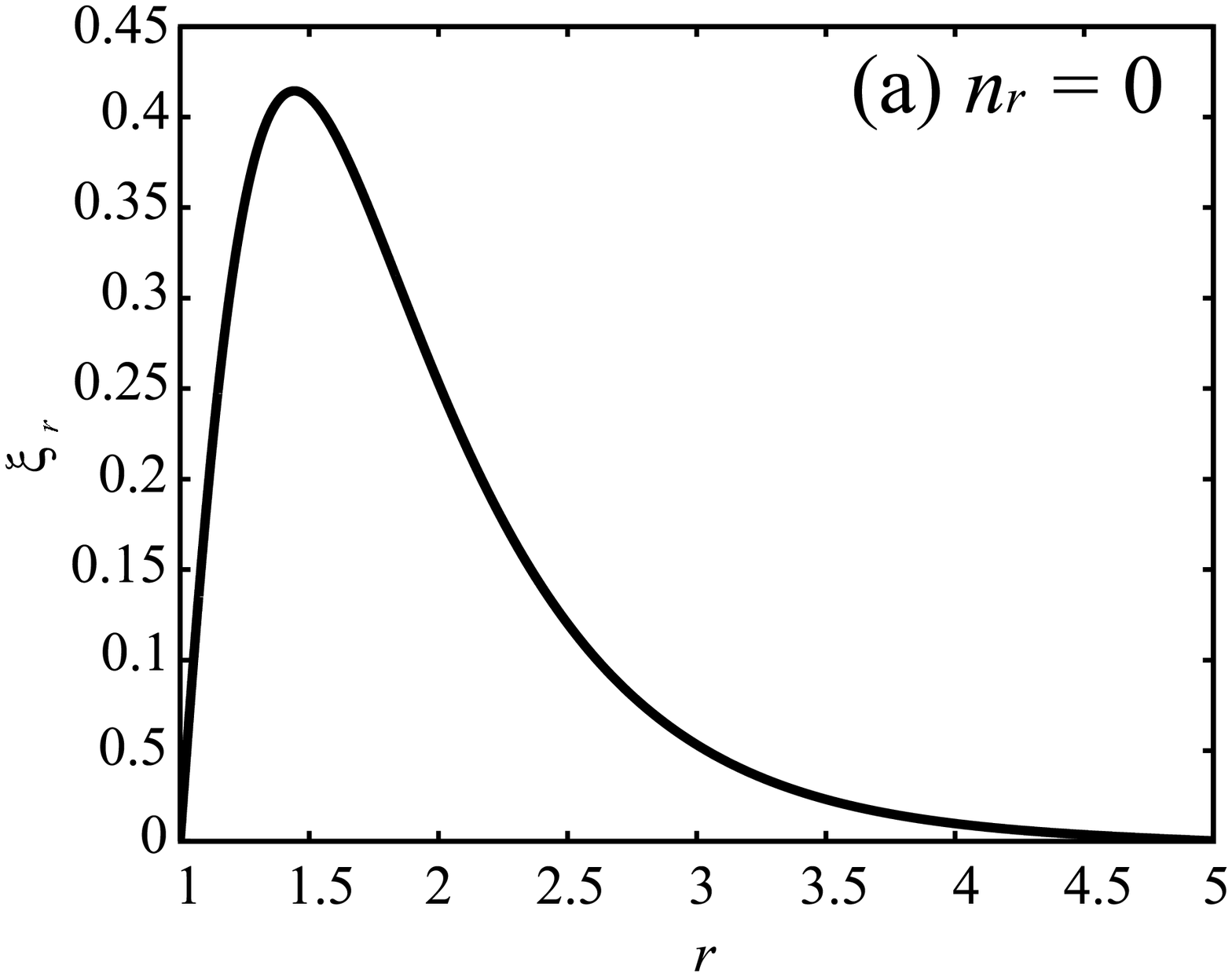}\\
  \includegraphics[scale=0.4]{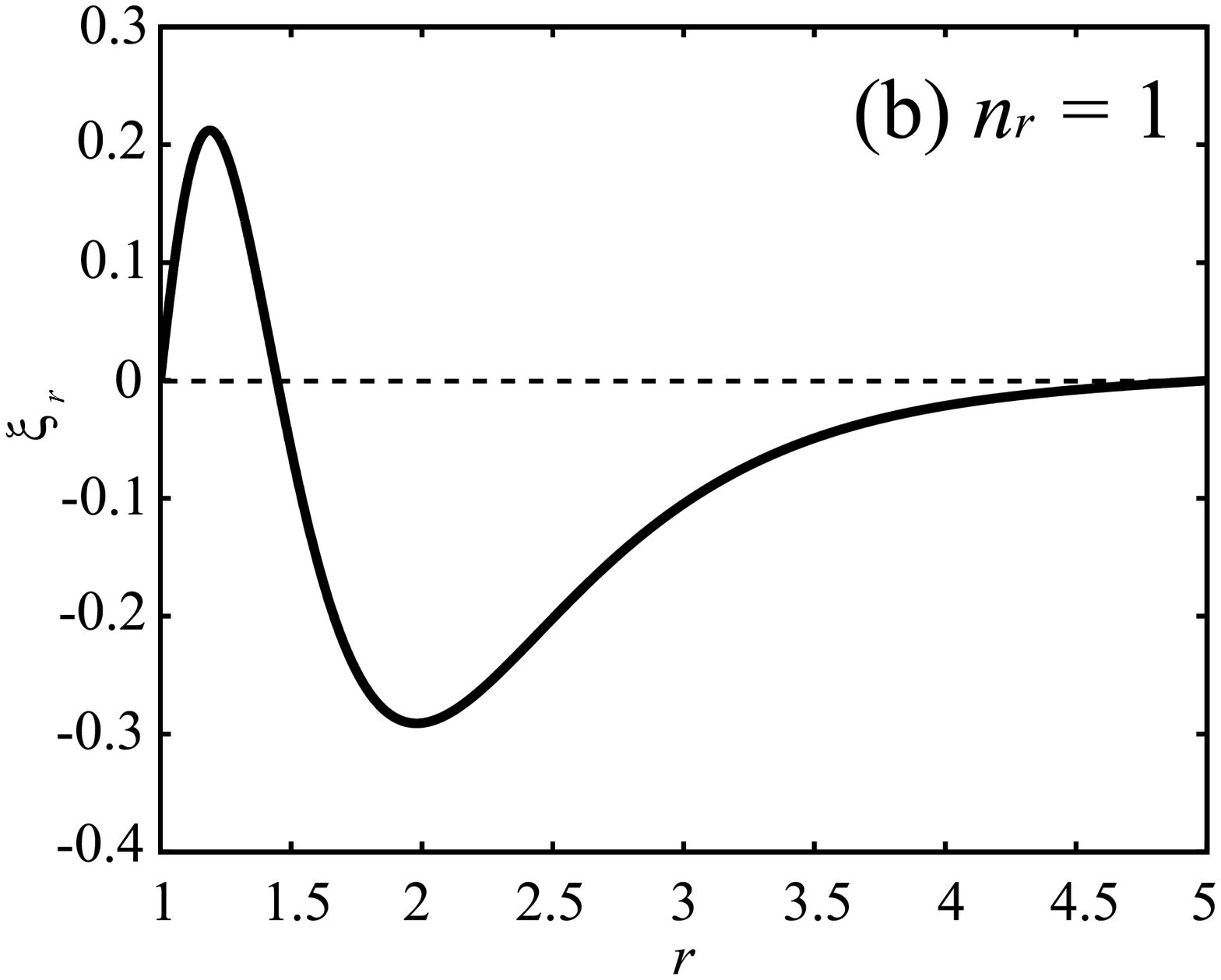}\\
  \includegraphics[scale=0.4]{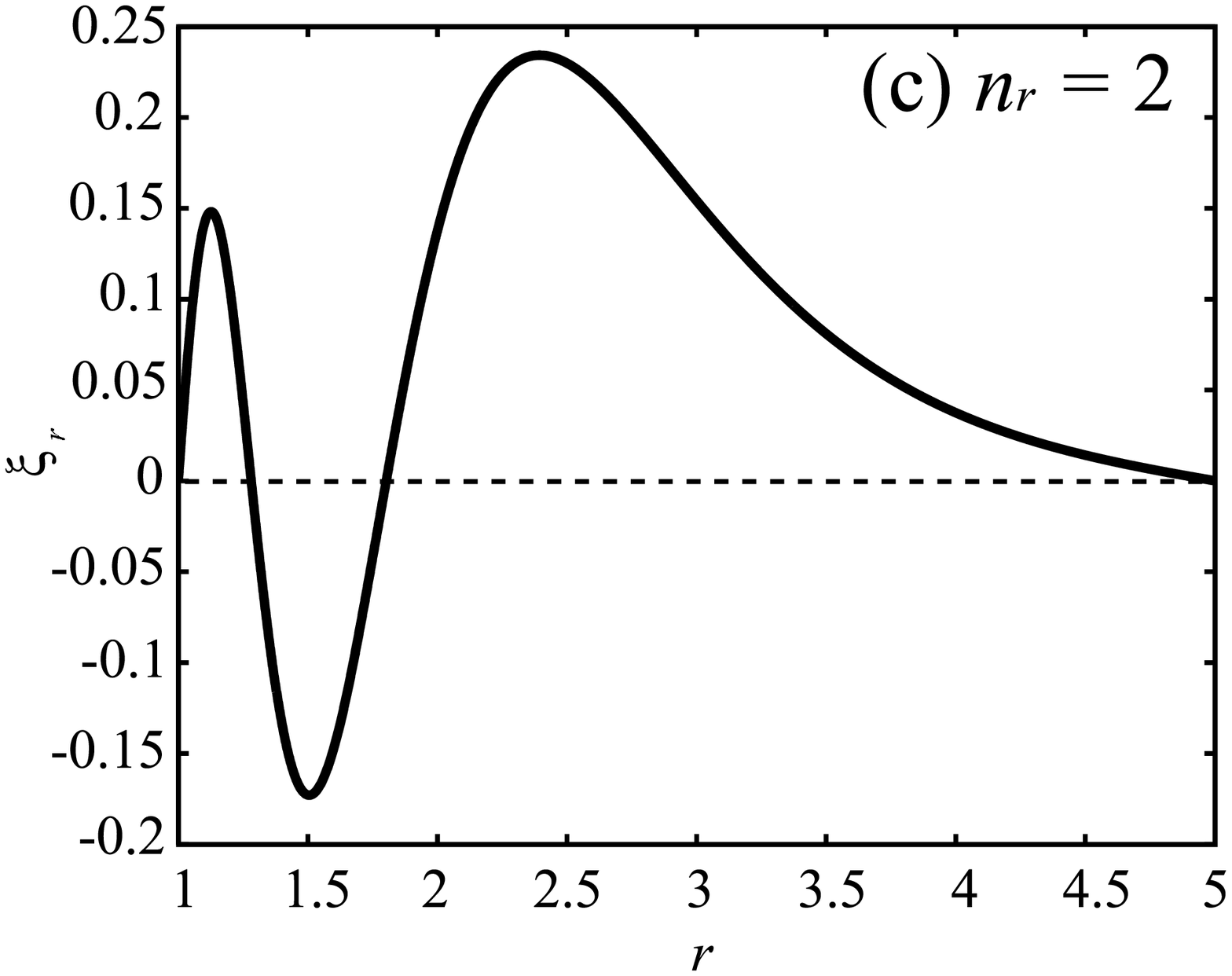}\\
  \caption{Eigen-functions in case $m=0$, $k=\pi/2$. Corresponding
  radial wave-numbers and eigen-frequencies are (a) $n_r=0$,
  $\omega=\pm0.2099i,~\pm0.5937$; (b) $n_r=1$,
  $\omega=\pm0.1908,~\pm0.5216$; (c) $n_r=2$,
  $\omega=\pm0.2625,~\pm0.4894$.\label{m=0,func}}
\end{center}
\end{figure}

\begin{figure}[p]
\begin{center}
  \includegraphics[scale=0.5]{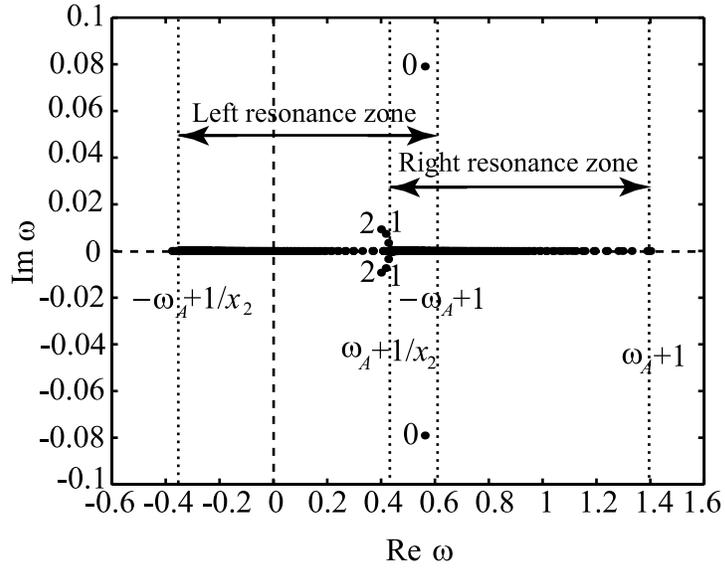}\\
  \caption{Eigen-frequency spectrum in case $m=1$, $k=\pi/2$. Dashed
  lines represent real and imaginary axes, dotted lines correspond to
  boundaries of Alfven resonance zones.
  Radial wave-numbers $n_r$ are shown for some unstable modes.}\label{m=1,spec}
\end{center}
\end{figure}

\begin{figure}[p]
\begin{center}
  \includegraphics[scale=0.4]{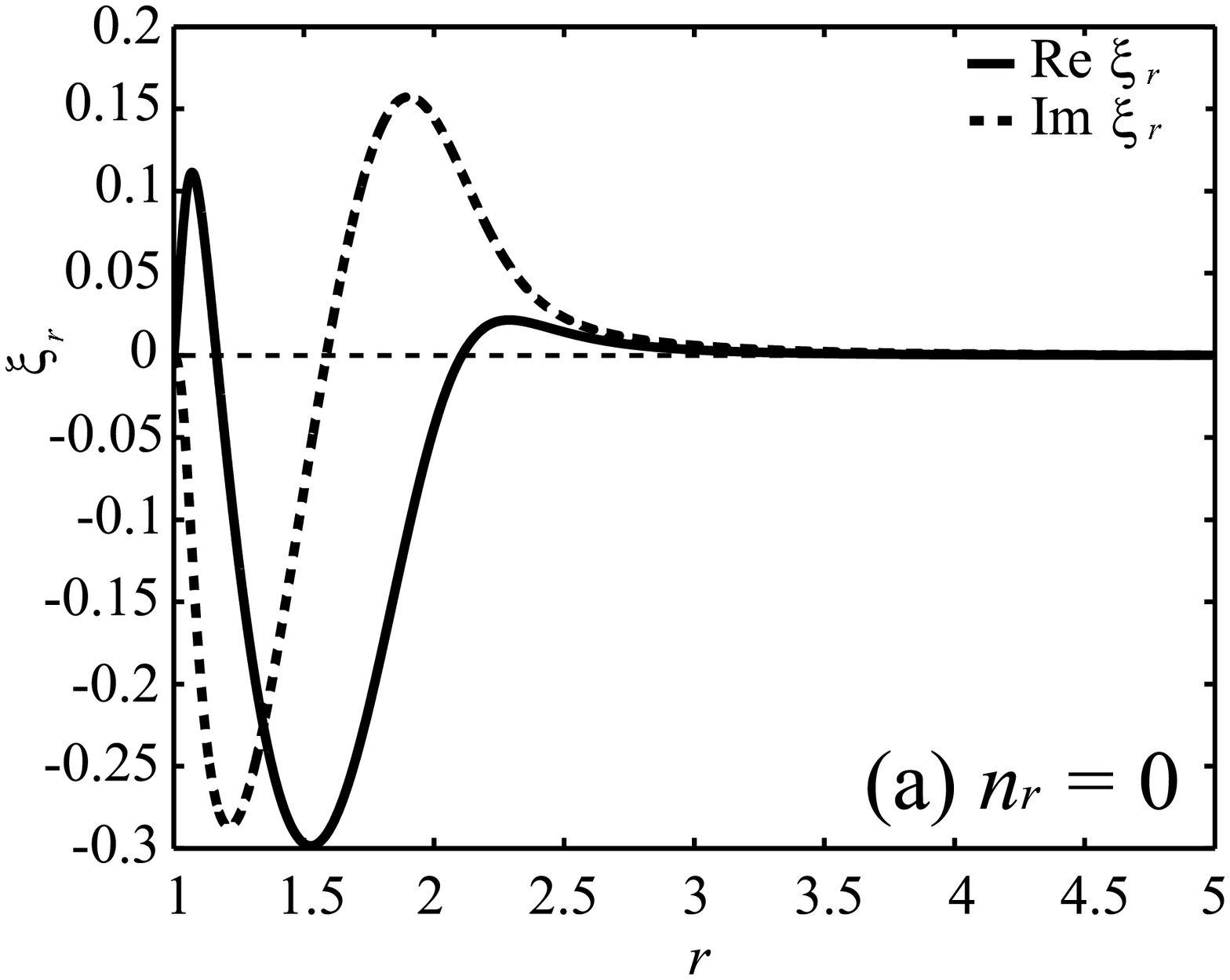}\\
  \includegraphics[scale=0.4]{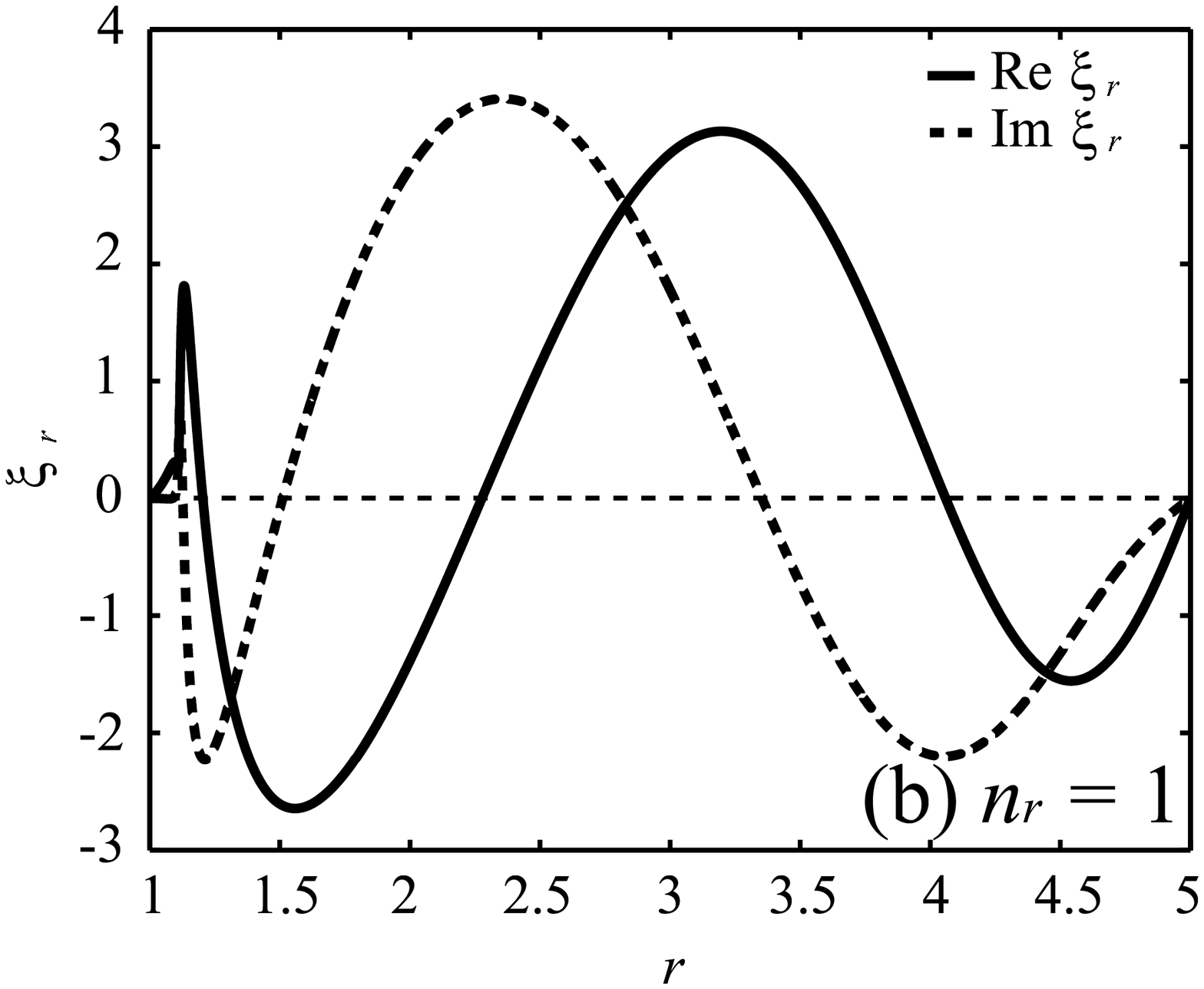}\\
  \includegraphics[scale=0.4]{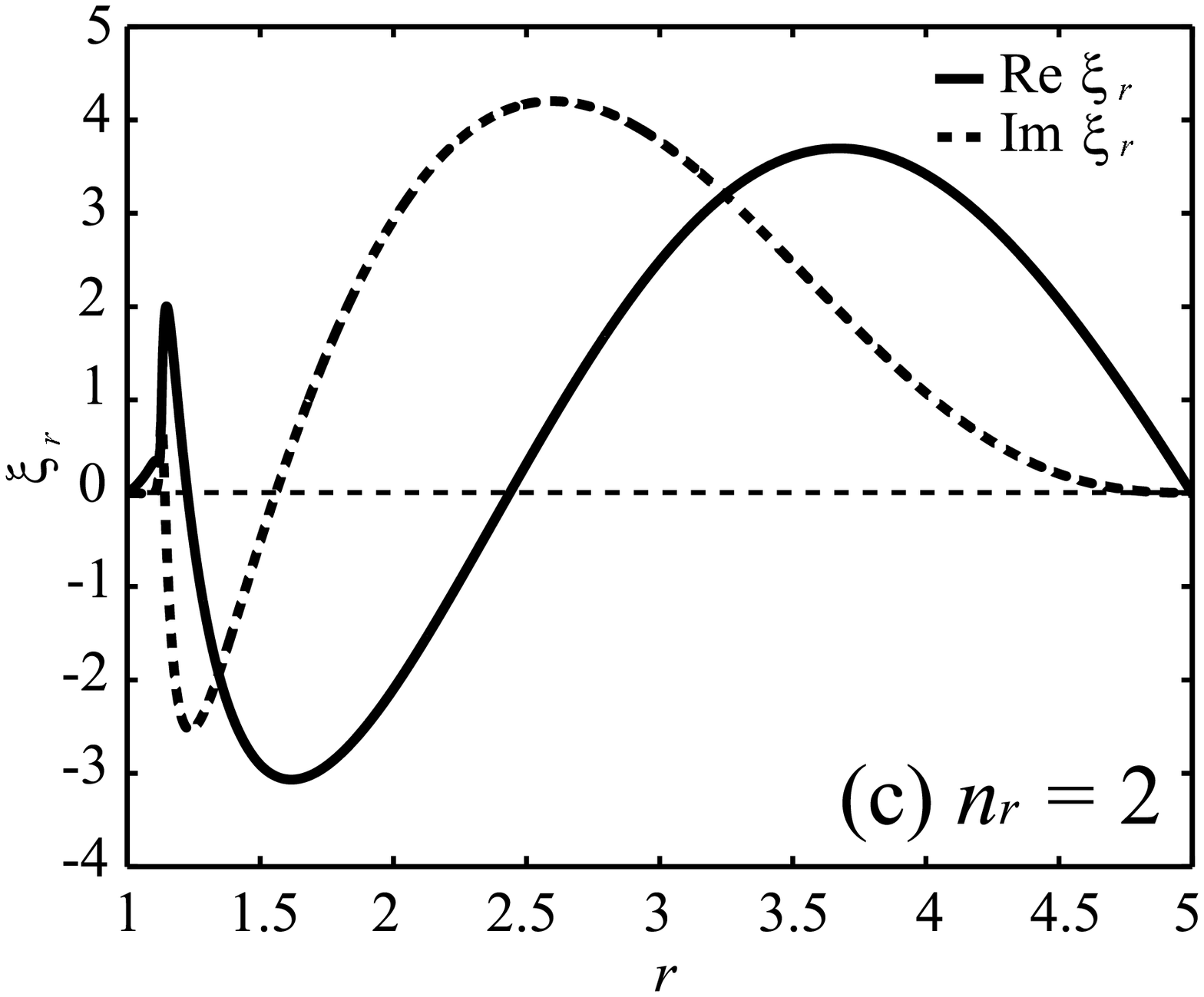}\\
  \caption{Eigen-functions in case $m=1$, $k=\pi/2$. Real
  (solid) and imaginary (dashed) parts are shown.
  Corresponding radial wave-numbers and eigen-frequencies are
  (a) $n_r=0$, $\omega=0.5623+0.0791i$; (b) $n_r=1$,
  $\omega=0.4173+0.0073i$; (c) $n_r=2$,
  $\omega=0.3994+0.0093i$. \label{m=1,func}}
\end{center}
\end{figure}

\begin{thebibliography}{200}

\bibitem{Vel} E. P. Velikhov, Sov. Phys. JETP \textbf{36}, 995 (1959).

\bibitem{Balbus} S. A. Balbus and J. F. Hawley, Astrophys. J. \textbf{376}, 214 (1991).

\bibitem{Rud1} G. R\"{u}diger and Y. Zhang, Astron. Astrophys. \textbf{378}, 302 (2001).

\bibitem{PPPL1} H. T. Ji, J. Goodman, and A. Kageyama, Mon. Not. R. Astron. Soc.
\textbf{325}, L1 (2001).

\bibitem{Los1} K. Noguchi, V. I. Pariev, S. A. Colgate, H. F. Beckley, and J.
Nordhaus, Astrophys. J. \textbf{575}, 1151 (2002).

\bibitem{Rud3} R. Hollerbach and G. R\"{u}diger, Phys. Rev. Lett. \textbf{95}, 124501
(2005).

\bibitem{Los2} K. Noguchi, V. I. Pariev, astro-ph/0309340 (2003).

\bibitem{Mariland} D. R. Sisan, N. Mujica, W. A. Tillotson, Y.-M. Huang, W. Dorland,
A. B. Hassam, T. M. Antonsen, and D. P. Lathrop, Phys. Rev. Lett.
\textbf{93}, 114502 (2004).

\bibitem{Holl} R. Hollerbach and A. Fournier, in
\emph{MHD Couette Flows: Experiments and Models}, edited by R.
Rosner, G. R\"{u}diger, and A. Bonanno, AIP Conf. Proc. No. 733
(AIP, New York, 2004), p. 114.

\bibitem{my} I. V. Khalzov and A. I. Smolyakov, Technical Physics \textbf{51}, 26 (2006).

\bibitem{Ostr} W.-T. Kim and E. C. Ostriker, Astrophys. J.
\textbf{540}, 372 (2000).

\bibitem{Curry1} C. Curry, R. E. Pudritz, and P. G. Sutherland,
Astrophys. J. \textbf{434}, 206 (1994).

\bibitem{Curry2} C. Curry and R. E. Pudritz, Mon. Not. R. Astron. Soc. \textbf{281}, 119 (1996).

\bibitem{Rud2} G. R\"{u}diger, M. Schultz, and D. Shalybkov, Phys. Rev. E \textbf{67},
046312 (2003).


\bibitem{PPPL3} J. Goodman and H. Ji, J. Fluid Mech. \textbf{462}, 365 (2002).

\bibitem{Frieman} E. A. Frieman and M. Rotenberg, Rev. Mod. Phys. \textbf{32}, 898 (1960).

\bibitem{Ogilvie} G. I. Ogilvie and J. E. Pringle, Mon. Not. R. Astron. Soc. \textbf{279}, 152 (1996).

\bibitem{Goed} R. Keppens, F. Casse, and J. P. Goedbloed, Astrophys. J. \textbf{569}, L121,
(2002).

\end{thebibliography}
\end{document}